\documentclass[12t]{article}

\usepackage{amsmath}
\usepackage{amssymb}
\usepackage{graphicx}

\def\S{{\cal{S}}}
\def\D{{\cal{D}}}
\def\H{{\cal{H}}}
\def\P{{\cal{P}}}
\def\A{{\cal{A}}}

\def\F{{\cal{F}}}

\def\Tr{{\rm Tr}}
\def\Sp{{\rm Sp}}
\def\Qm{{\hat{\Pi}}}
\def\Qv{{\hat{V}}}
\def\Qi{{\hat{I}}}
\def\Qa{{\hat{a}}}
\def\ra{{\rangle}}
\def\la{{\langle}}
\def\C{{\mathbb{C}}}
\def\Z{{\mathbb{Z}}}
\def\Ac{{I_{{\rm ac},1}}}

\title{Information capacity formula of quantum optical channels}
\author{
M. Sohma\\
{\it \small Matsushita Research Institute  Tokyo, Inc., Japan}\\
O. Hirota\\
{\it \small Research Center for Quantum Communication,}\\
{\it \small Tamagawa University, Machida,
Tokyo, Japan.}
\\{\it \small  and}\\{\it \small  CREST, JST: Japan  Science and Technology}  }
\date{\empty}

\begin{document}
\maketitle
\begin{abstract}
The applications of the general formulae of channel capacity developed in
the quantum information theory to evaluation of information transmission
capacity of optical channel are interesting subjects.
In this review paper,  we will point out that the formulation based on only
classical-quantum channel mapping model may be inadequate when one takes
into account a power constraint for noisy channel.
To define the power constraint well,
we should explicitly consider how quantum states are conveyed 
through a transmission channel.
Basing on such consideration, 
we calculate a capacity formula for an attenuated noisy optical channel
 with genreral Gaussian state input; this gives
certain progress beyond the example in our former paper \cite{Holevo:99}.
\end{abstract}

\section{Introduction}

During about twenty years, the information theory for quantum
channel has been devised\cite{Bennett:98}. It is called quantum information theory.
The most fundamental result in it is the formula of channel capacity,
which might be
an interesting subject for optical communication issue;
if one wants to know the ultimate capability of information transmission,
one has to know the channel capacity. 
%Hence such a theory might be
%interesting subject for optical communication issue.

Here we introduce short survey of a 
channel model and its capacity theory \cite{Hirota:00}.
Let ${\{\vert\alpha_m\rangle}\}$, $m=1, 2$ be binary letter states,
whose state overlap is
$\kappa=\langle\alpha_1\vert\alpha_2\rangle$. By $n$-th
extension,
we choose $M$ codeword states $\{ \vert \psi_1\rangle, \cdots,
\vert \psi_M\rangle \}$ $(M\le2^n)$ from the $2^n$ possible sequences
of length $n$, and use them with
respective {\it a priori} probabilities $\{\pi_1, \cdots, \pi_M \}$, where
$\vert \psi_i\rangle = \vert\alpha_{m_1}(i)\rangle\otimes\vert\alpha_{m_2}(i) 
\rangle\
\vert\otimes\alpha_{m_3}(i)\rangle\otimes\cdots\otimes
\vert\alpha_{m_n}(i) \rangle$.
We can regard a codeword state as one signal state.
Since the codeword states are linearly
independent, they span the $M$ dimensional Hilbert space. Then the optimization
problem of quantum decoding is reduced to the detection problem for $M$-ary no
coded quantum signals.
The above statements are the definition of the quantum coding and
decoding for Shannon information.
Here, the decoding means abstract representation of optical receivers,
which are described by so called positive operator valued measure(POVM).
The decoding processes can be classified into two classes in general.
One of them is decoding based on {\it "individual separate measurement"}.
It means that each bit of codeword is measured and a received codeword is
decoded by the data processing based on $n$ measured variables.
However, the super additivity is not allowed by decoding based on the
individual
separate measurement\cite{Sasaki:98}.
This decoding operator (or detection operator)
is described as follows:
\begin{equation}\label{indiv}
\hat{\mit\Pi}_i=\otimes_{k=1}^n\hat{{\mit\Pi}_k}^{(i)}
\end{equation}
where
$\hat{{\mit\Pi}_k}^{(i)}$ is the optimum detection operator for letter states.
The other is
decoding based on  {\it entangled} or {\it collective measurement},
which is called {\it "entangled decoding"}.
In the latter case, the codeword state is regarded as single state and
a set of $M$ detection operators for  $M$-ary no
coded quantum states is applied to decide them. Therefore, one can obtain only
$M$ values of detection for $M$ codewords in the decoding process.
In this model, there is a bound for capacity, so called
Holevo bound.
His general formula involves also the mixed state case.
%that the transmitter states
%are mixed state. 
The detailed discussion is given in the later sections.
Then, for pure state case, Hausladen, Jozsa, Schumacher, Westmoreland,
and Wooters\cite{Haus:96} proved that the Holevo bound for the information
transmission by pure state is really channel capacity.
That is, the zero error channel capacity $C$ is defined by
$\lim_{n\to\infty}{C_n}/{n}$ and it is given by the
maximization of von Neumann entropy with respect to {\it a priori}
probability for an ensemble of signal states.

Recently, Holevo\cite{Holevo:98}, and Schumacher-Westmoreland\cite{Schu:97} proved that
the Holevo bound  is also the channel capacity in the general case (mixed state case).
Finally, Holevo gave the formula for the continuous alphabet\cite{Holevo:98b}.
As a result, the capability of information transmission increases by design of
coding and decoding for $n$th extension.
Then a fundamental property, super additivity 
of channel capacity by extended codes, ${C}_{n}$+ ${C}_{m}$ $\le$ ${C}_{n+m}$, was found.
%
%Such a  fundamental property  like ${C}_{n}$+ ${C}_{m}$
%$\le$ ${C}_{\rm n+m}$
%comes
%from a property so called super additivity of channel capacity
%by extended codes.

In this paper, we survey the general scheme of the theory of channel
capacity, and show how to apply them to optical communication process.
Finally, the general capacity formula for optical channel with energy loss
and back ground noise is given, which corresponds to Shannon formula: $C =
\log(1+{S}/{N})$.

In Section 2, we present a quantum model of optical communication, 
"{\it attenuated noisy channel}".
In preparation for calculating the capacity of such a channel,
we survey the results about the channel coding theorem in Sections 3 and 4:
In Section 3, we consider the "{\it classical-quantum channel model}", 
which is a generalization of attenuated noisy channel, and give the 
general formula
of its capacity. To calculate the channel capacity, we need to solve 
an optimization problem included in the general capacity formula 
(see the formulation (\ref{cont def})). In Section 4 we calculate
the channel capacity for quantum Gaussian channel, 
solving the optimization problem.
Modifying this calculation, we obtain the capacity of 
the attenuated noisy channel in Section 5, where
the formulae (\ref{cap1form}) and (\ref{cap3form}) are main results
of this paper.
In Section 6 
we discuss a discretization of quantum continuous channel.

\section{Quantum Model of Optical Communication}\label{qmooc section}
We start with presenting a quantum model of optical communication,
which is the main subject of our interest.
In our model, one mode Bosonic states, which the transmitter outputs, 
are conveyed through an attenuation channel with classical noise.
We can formulate this model as follows.
\newline {\bf (i)  transmitted Bosonic states}
\newline
Let $\Qa$ be an annihilation operator on a Hilbert space $\H$.
%Then the one mode squeezed state is represented as
%$\S(\zeta)|0\rangle$ 
%with $\zeta=\gamma {\rm e}^{i\theta}$ and 
%$\S(\zeta)=\exp[ (\zeta^{*}\Qa^2-\zeta(\Qa^{\dagger})^2)]$.
We take as an input alphabet $\A$ the complex plane $\C$ 
or nonnegative integer $\Z_0^{+}$,  and assume that
the transmitter sends 
a  Bosonic state 
$\tilde{\rho}(\mu)$ corresponding to a letter $\mu \in \A$.
In particular our concern is concentrated on two examples:
(a)
$\tilde{\rho}(\mu)=\D(\mu)\tilde{\rho(0)}$
with  a squeezed state $\tilde{\rho}(0)$ and 
the unitary {\it displacement} operator  
$\D(\mu)=\exp (\mu \Qa^{\dagger}-\bar{\mu}\Qa)$, 
(b) $\tilde{\rho}(n)$ are number states $|n\ra\la n |$ ($n\in\Z_{0}^{+}$).
\newline {\bf (ii) linear attenuator with classical noise}
\newline
The linear attenuator 
with coefficient $k\leq 1$ and Gaussian noise 
with variance $N_c$ is described by the transformation 
\begin{equation}\label{attenuator}
\Qa'=k\Qa+\sqrt{1-k^2}\Qa_0+\xi,
\end{equation}
in the Heisenberg picture.
In the equation (\ref{attenuator}), $\Qa_0$ is an annihilation operator 
in another mode in the Hilbert space $\H_0$ of an "environment" and 
$\xi$ is a complex random variable with zero mean and variance $N_c$.
We assume that the environment is initially in the vacuum state.
We denote by $\Gamma$ the corresponding 
transformation of states in the Shr\"{o}dinger picture:
$\Tr \tilde{\rho}(\mu) \Qa'=\Tr \Gamma[\tilde{\rho}(\mu)] \Qa$.
\newline {\bf (iii) detection process}
\newline
Consider the {\it code of size} $M$, $\{ u_1, u_2, ... ,u_M \}$, consisting of 
codewords of length $n$, $u_i=(\mu_{i_1},...,\mu_{i_n})$, where 
$\mu_{i_j}$ is selected from the continuous input alphabet $\A=\C$ or 
the discrete  one $\A=\Z_0^{+}$.
Then the codeword $u_i$ is related 
to a product state $\rho(\mu_{i_1})\otimes\cdots\otimes \rho(\mu_{i_n})\in 
\H\otimes\cdots\otimes\H\equiv\H^{\otimes n}$, where 
$\rho(\mu_{i_j})=T[\tilde{\rho}(\mu_{i_j})]$.
This correspondence gives a channel mapping stated in Section \ref{ct sec}.
A detection process is given by a detection operator $\Qm_i$,
which is a positive operator-valued measure (POVM) on $\H^{\otimes n}$
\cite{Holevo:73} defined as
\begin{equation}
\begin{split}
\Qm_i & \geq 0, \quad \forall i, \\
\sum_{i=1}^{M}\Qm_i&=\Qi.
\end{split}
\end{equation}
The POVM represents a measurement process and decision 
for a signal to be $u_i$ based on the measurement result.
Using POVM, the conditional probability of
the output $u_j$, 
given the input was $u_i$, is obtained as,
\begin{equation}\label{error1}
P(u_j|u_i)=\Tr \rho_{u_i} \Qm_j.
\end{equation}

How to define information quantities for such a channel, is 
shown in the next section using a more general formulation.

\section{Coding Theorem of Quantum Channel for Shannon Information}\label{ct sec}
In this section we survey the coding theorem for classical-quantum channels,
where Shannon information is conveyed by quantum states.
%In particular, a continuous classical-quantum channel given in  
%Section \ref{ias section} is a generalization of the model presented in 
%the previous section.
\subsection{Finite alphabet system}\label{sec cap}
\subsubsection{Holevo-Schumacher-Westmoreland theorem}\label{sec cap2}
Let $\H$ be a Hilbert space. The classical-quantum channel
(coding channel) with discrete alphabet $\A=\{1,..,a\}$  consists of the 
mapping $i\to \rho_i$ from the input alphabet to the set 
of density operators in $\H$. The input is described by an 
{\it a priori} probability distribution $\pi=\{\pi_i\}$ on $\A$.
A quantum detection process (decoding channel) is described 
by POVM $\Qm=\{ \Qm_j \}$ on $\H$. 
Like Eq. (\ref{error1})
 the conditional probability of
the output $j$, 
given the input was $i$, is obtained as,
\begin{equation}\label{error2}
P(j|i)=\Tr \rho_i \Qm_j,
\end{equation}
and the Shannon's mutual information is given by
\begin{equation}\label{m form1}
I_1(\pi,\Qm)=\sum_j\sum_i \pi_i P(j|i)\log \left( 
\frac{P(j|i)}{\sum_{k}\pi_k P(j|k)}
\right ).
\end{equation}
Moreover let us consider the product channel in the tensor 
product Hilbert space $\H^{\otimes n}=\H\otimes \cdots\otimes \H$ 
with the input alphabet $\A^n$ consisting of words $u=(i_1,...,i_n)$ 
of length $n$, with density operator 
\begin{equation}
\rho_u=\rho_{i_1}\otimes \cdots \otimes \rho_{i_n}.
\end{equation}
If $\pi$ is a probability distribution on $\A^n$ and $\Qm$
is a POVM on  $\H^{\otimes n}$, 
we can define the information quantity $I_n(\pi,\Qm)$ by a formula similar to 
(\ref{m form1}).
Now let us define 
\begin{equation}
C_n=\sup_{\pi,\Qm}I_n(\pi,\Qm).
\end{equation}
Then 
we have the property of super additivity 
\begin{equation}
C_n+C_m\leq C_{n+m},
\end{equation} 
and hence the following limit exists:
\begin{equation}\label{m form1 end}
C=\lim_{n\to \infty}C_n/n.
\end{equation}
This limit gives a definition of {\it  capacity}
 of the initial channel.
Here it should be emphasized that in the  definition of this quantity
we employ an {\it entangled measurement}.
%which describe simultaneously the quantum
%measurement process and decision process, that is, 
%which outputs directly the decision symbols
%corresponding to the hypothesis for input quantum states.
The use of such measurement causes the superadditivity, 
which is characteristic of quantum system.
In contrast, in a semi classical case, 
we consider {\it individual separate measurement} 
of the form (\ref{indiv}),
%which is described as
%\begin{equation}
%\Qm_i=\otimes_{k=1}^n \Qm_k^{(i)},
%\end{equation}
and carry out error correction based on data processing of measured value.
Such a detection strategy 
never produces the super additivity, 
but only achieves $C_1$. 

Using the von Neumann entropy, which is defined as $H(\rho)=-\Tr\rho \log \rho$ for
a density operator $\rho$, we can obtain 
a simple formula of 
the capacity:
\newline {\bf Theorem}\cite{Holevo:98,Schu:97}:
\newline
{\it The capacity of  arbitrary 
signal states $\rho_i$, having finite entropy $H(\rho_i)$, 
is given by
\begin{equation}\label{Hol bound}
C=\lim_{n\to\infty} {C_n}/{n}= \max_{\pi} \Delta H(\pi),
\end{equation}
where 
\begin{equation}
\Delta H(\pi)=
 H\left( \sum _{i\in \A} \pi_i \rho_i \right)-
\sum_{i\in \A}\pi_iH(\rho_i).
\end{equation}
}

In general the computation of the quantity $C$ is very difficult, 
but, fortunately, if the signal states $\rho_i$ have a certain symmetry, 
the analytical solution can be obtained:
Let us consider signal states given as
\begin{equation}
\rho_i=\Qv^{i-1}\rho_1\Qv^{\dagger i-1}
\end{equation}
where $\Qv^{\dagger}\Qv=\Qv\Qv^{\dagger}=\Qv^M=\Qi$.
Then it is shown \cite{Kato:99b} that the capacity is achieved by a 
uniform distribution on {\it a priori} probabilities.
In particular, when $\rho_1$ is a pure state represented as $\rho_1=|\psi \ra\la \psi|$,
the capacity is calculated as,
\begin{equation}
C=-\sum_{j=1}^M\lambda_j\log\lambda_j,
\end{equation}
where $\lambda_j$ is
\begin{equation}
\lambda_j=\frac{1}{M}\sum_{k=1}^M\la\psi|\Qv^{k-1}|\psi\ra\exp
\left( -\frac{2j(k-1)\pi i}{M} \right).
\end{equation}

%The entropy bound shown in \cite{Holevo:71} leads to the inequality
%$C\leq \max_{\pi}\Delta H(\pi)$, and  hence, for  proving the theorem, 
%it suffices to show the converse inequality.
%To do this, we estimate the error probability 
%with a semi optimal quantum measurement.
\subsubsection{$C_1$ and Accessible information}
As described in Sec. \ref{sec cap2},
 most essential property of quantum theory for
capacity  is
the super additivity. For research of coding scheme achieving the capacity,
we have to clarify properties of  the super additivity. As a first step,
 we consider {\it accessible information}, 
which is defined to be
\begin{equation}
I_{{\rm ac},n}(\pi)=\sup_{\Qm}I_n(\pi,\Qm),
\end{equation}
for fixed {\it a priori} distribution $\pi$.
A necessary condition for quantum measurement $\Qm$
achieving the accessible information $\Ac(\pi)$
was given as follows\cite{Holevo:73}:
\begin{equation}
\hat{\mit\Pi}_i \left(
\F_i - \F_j
\right)\hat{\mit\Pi}_j =0,
\forall i,j,
\label{eqn:infoopt}
\end{equation}
where
\begin{equation}
\F_j=
\displaystyle\sum_l \pi_l \log \left[\frac{P(j|l)}{\sum_k
\pi_k P(j|k)}\right]
{\rho}_l.
\label{eqn:infooptFj}
\end{equation}
Since the above equation provides only a necessary
condition for detection operators, we cannot, in general,
solve the problems.
From another point of view, Davies proved theorems\cite{Davies:78} of 
conditions for optimum detection operators which give accessible
information. His theorems require that one has to take into
account a number of detection operator $J$ corresponding to a
size of detection scheme to get an accessible information.
That is, $d\leq J\leq d^2$, where $d$ is the dimension of
Hilbert space of letter states.

Now let us consider $C_1$.
Although to derive $C_1$ is very difficult,
Davies's theorems play an important role to search it
actually.
%the capacity of letter states, $C_1$. 
Based on these
theories, Levitin\cite{Levitin:95}, Fuchs and Caves\cite{Fuchs:94}, and
Fuchs and Peres\cite{Fuchs:96} gave
some examples of $C_1$ as almost final result, and they
clarified an importance of this problem in a quantum
information theory. On the other hand, Ban\cite{Ban:97} 
and Osaki\cite{Osaki:98}
proved that if the signal states are group covariant, then the
square root measurements or minimax solutions in quantum
detection theory satisfy Holevo's necessary condition for
information criterion, and finally Osaki numerically\cite{Osaki:98} and
Ban analytically \cite{Ban:99} derived ${C}_{\rm 1}$ of the binary pure
states taking all parameters into account.
The final result: ${C}_{\rm 1}$ 
is so simple as follows:
\begin{equation}
C_1 = 1 - H(P_{\rm e}(2))
\end{equation}
where $H(P_{\rm e}(2))$ is entropy for error probability,
$P_{\rm e}(2)$ is the minimum error probability in the 
detection problem for binary states,
\begin{equation}
P_{\rm e}(2) = \left[1 - (1-{\kappa}^2)^{\frac{1}{2}}\right]/2
\end{equation}
where $\kappa$ is the inner product between two pure states.

Let us turn the problem into the system with linearly 
dependent state. If there are three states in $2$ dimensional Hilbert 
space, then they are linearly dependent. When they have equal 
angle each other, the channel capacity of letter states 
is $C_1 = 0.6698$, where
$C_1$ is given by
$\pi_1=\pi_2 = 1/2$, and $\pi_3 = 0$.
However, we have no exact solution for other cases.

\subsection{Infinite alphabet system}\label{ias section}
Let us turn to the infinite alphabet system;
we present a general formula of capacity for 
a quantum continuous channel according to
\cite{Holevo:99}. 

Take as the input alphabet ${\A}$ an arbitrary Borel
subset in a finite-dimensional Euclidean space ${\cal E}$.
Then the quantum continuous channel is described by a
weakly continuous mapping ${\mu\to \rho(\mu)}$ from 
the input alphabet $\A$ to the set of density operators 
in ${\cal H}$,
where we assume that the states $\rho(\mu)$ have a finite 
von Neumann entropy $H(\rho(\mu))=-\Tr \rho(\mu) \log \rho(\mu)$, moreover
\[
\sum_{\mu\in \A }H(\rho(\mu))< \infty.
\]

Like the discrete case, we treat the product memoryless channel in the 
Hilbert space ${\cal H}^{\otimes n}={\cal H}\otimes\cdots\otimes{\cal H}$ ($n$ copies).
Then the signal $u_i=(\mu_{i_1},...,\mu_{i_n})$ 
is related to a density operator
$\rho(\mu_{i_1})\otimes\cdots\otimes \rho(\mu_{i_n})$.
In a similar way as the classical case,
we impose the power constraint 
\begin{equation}\label{ic0}
f(\mu_{i_1})+\cdots +f(\mu_{i_n})\leq nE
\end{equation}
on the signals $u_i=(\mu_{i_1},...,\mu_{i_n})$, 
%of the classical-quantum channel
%\begin{equation}\label{c-q map}
%$u_i\to \rho(\mu_{i_1})\otimes\cdots\otimes \rho(\mu_{i_n})$,
%\end{equation} 
where $f$ is a continuous positive function on ${\cal E}$.
In particular, we denote by $\P_1$ a set of probability distributions $\pi$
on $\A$ satisfying
\begin{equation}\label{ic1}
\int _{\A} f(x)\pi(dx)\leq E.
\end{equation}
Without the constraint we cannot obtain anything but meaningless results, that is, 
arbitrary high transmission rates can be achieved with arbitrary low error
probability with essentially no coding.

%This formulation is a generalization of the attenuated noisy channel 
%given in Sec. \ref{qmooc section}.
%It should be emphasized that the density operator 
%$T[\tilde{\rho}(\mu)]$ in the model of the attenuated noisy channel
%corresponds to $\rho(\mu)$ in the general formulation described above.
%In this general formulation only received signal states, that is
%inputs for detection process, are considered, and 
%effects of other elements, 
%signal states sent from the transmitter, attenuator, 
%and classical noise, are ignored.

The theory of capacity of continuous channel was established 
by Holevo. Let us sketch the outline of the theory according to \cite{Holevo:98b}.
We consider discretization of the channel 
by taking a priori distributions with discrete support
\begin{eqnarray}\label{discrete support}
\pi (d\mu)=\sum_i \pi_i \delta_{\mu_i}(d\mu), \label{discrete prob}
\end{eqnarray}
where $\{ \mu_i \}$ is arbitrary countable collection of points and 
\begin{eqnarray}
\delta_{\mu}(B)\equiv
\left\{
\begin {array}{rl}
1 & \mbox{if $\mu \in B$,}\\
0 & \mbox{if $\mu \notin B$.}
\end{array}
\right . 
\end{eqnarray}
Then the capacity $C$ is defined as follows.
\begin{subequations}\label{cont capacity sub}
\begin{eqnarray}
C&=&\sup_{\pi \in  \P'_1}\Delta H(\pi ),\label{cont capacity}\\
\Delta H(\pi) &\equiv & H \left ( \int _\A \rho_{x} \pi (dx) \right )  - \int_\A  H(\rho_{x}) \pi(dx),
\end{eqnarray}
\end{subequations}
where $\P'_1$ is a set of {\it  a priori} distributions with discrete support 
satisfying input constraint (\ref{ic1}).
Holevo proved \cite{Holevo:98b} that all rates below the channel capacity 
are achievable, 
and that under some assumptions
the channel capacity $C$ is equal to 
\begin{eqnarray}
\sup_{\pi \in \P_1}\Delta H(\pi ).\label{cont def}
\end{eqnarray}

So far we have described the theory for channel capacity.
As a result, the channel coding theorem has been proved for quantum
channel which corresponds to quantum measurement process.
This theorem provides interesting results. That is, the channel capacity is
exactly equal to the Holevo bound which is defined without the model of
quantum measurement process, while the capacity is defined for
channel model of the quantum measurement process as described in 
the equations (\ref{m form1})-(\ref{m form1 end}).
So if one wants to calculate only channel capacity, one needs not
the physical channel model.
In other word, the capacity is given only by the ensemble of prepared
states
or density operators in front of the measurement. Thus, sometimes,
in mathematical papers the channel is defined by a mapping from
classical alphabet to quantum states or density operators;
it is called classical-quantum channel. 
However, in order to evaluate the capacity of the attenuated noisy channel
described in Section \ref{qmooc section},  we should consider 
a slightly different channel model, which is described in the next subsection.
%It may work very well when one considers the problem of channel capacity.
%So, in Section \ref{sec4} we employ such a channel model.
%But we will need certain modification.
%Readers should not confuse the concept of such a channel. In physical
%sense, channel is
%attenuation process, measurement process, and so on.

\subsection{Channel model for describing a power constraint}\label{a2}
\begin{figure*}[t]　
\center
\includegraphics[scale=0.6,angle=0]{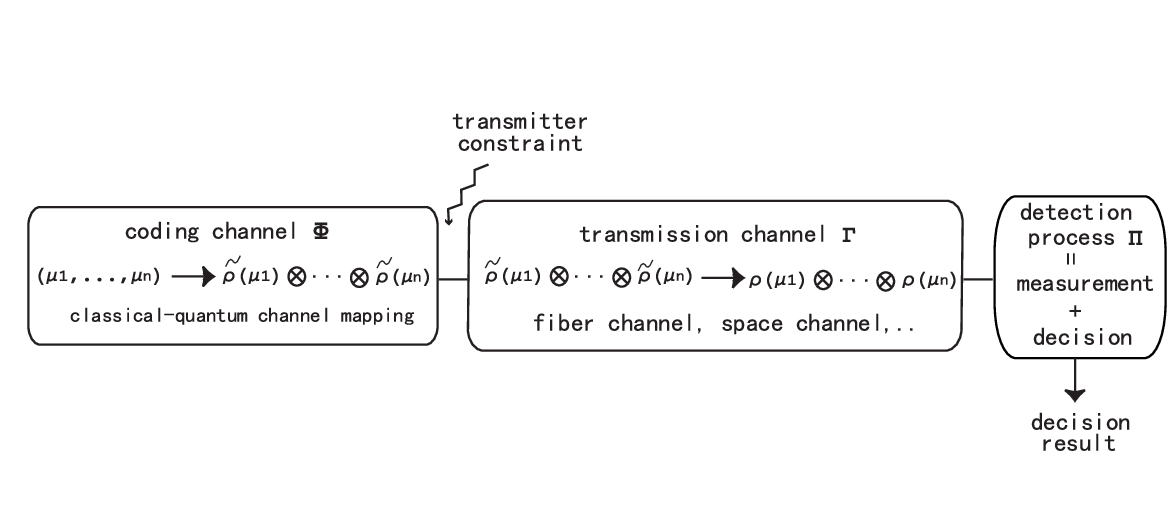}
\caption{channel model with transmitter constraint}\label{ch model}
\end{figure*}
We present channel model consisting of three components as shown in Figure \ref{ch model}:
\newline
(1) coding channel:
In the coding channel codewords $(\mu_1,...,\mu_n)$ are related to 
quantum states $\tilde{\rho}(\mu_1)\otimes\cdots\otimes\tilde{\rho}(\mu_n)$ 
respectively. Thus the coding channel is described as classical-quantum 
channel mapping $\Phi:(\mu_1,...,\mu_n)\to 
\tilde{\rho}(\mu_1)\otimes\cdots\otimes\tilde{\rho}(\mu_n)$.
\newline
(2) transmission channel:
Through the transmission channel
quantum states $\tilde{\rho}(\mu_1)\otimes\cdots\otimes\tilde{\rho}(\mu_n)$
are conveyed, and transformed to quantum states $\rho(\mu_1)\otimes\cdots
\otimes\rho(\mu_n)$ respectively. 
Thus the transmission channel is described by 
completely positive map $\Gamma: 
\tilde{\rho}(\mu_1)\otimes\cdots\otimes\tilde{\rho}(\mu_n)\to
\rho(\mu_1)\otimes\cdots
\otimes\rho(\mu_n)$.
(3) detection process.
\newline
This model gives a generalization 
of the attenuated noisy channel.

When we consider only (discrete) channel without power constraint,
we can apply the general capacity formula (\ref{Hol bound}) directly to
the above model by regarding $\bar{\Phi}^{(n)}=\Gamma^{(n)}\circ\Phi^{(n)}$ 
as the classical-quantum channel mapping.
On the other hand, for a continuous channel,
we should consider a constraint on the 
average power of signals
 $\tilde{\rho}(\mu_1)\otimes\cdots\otimes\tilde{\rho}(\mu_n)$ 
 that the transmitter outputs; such a constraint is called {\it transmitter}
 constraint in the following. Then we have no way to describe the power 
constraint function $f$ in (\ref{ic0}) and (\ref{ic1}) by using only $\bar{\Phi}^{(n)}$.
In other words we cannot formulate the optimization in (\ref{cont capacity sub}) 
only by $\bar{\Phi}^{(0)}$, while we can represent $\Delta H(\pi)$ 
as a function of $\pi$ and $\bar{\Phi}^{(0)}$.
This is the reason why we should explicitly distinguish 
coding channel $\Phi^{(n)}$ from transmission channel $\Gamma^{(n)}$ 
in our model.
Although use of the power constraint on the energy of 
input signals $(\mu_1,...,\mu_n)$ does not produce 
such a problem, it is not suitable for 
evaluation of the capacity of the attenuated noisy channel.

In the attenuated noisy channel the transmission map 
$\Gamma^{(n)}$ is parametrized by $k$ and $N_c$, which are defined in 
Section \ref{qmooc section}. As a result, its channel capacity is given as a function 
of $k$, $N_c$, and $\Phi^{(0)}(0)=\tilde{\rho}(0)$ in Section \ref{sec5}.

\section{Channel Capacity Formula for quantum Gaussian States}%*
\label{sec4}
%                                                             *
%**************************************************************
In this section we calcuate the channel capacity, 
based on the channel mapping model $\mu\to\rho(\mu)$
with {\it input} constraint according to \cite{Holevo:99}.
In particular we treat the case where $\rho(\mu)$ are quantum 
Gaussaian states.
The results will be applied to calculation of capacity 
for the attenuated noisy channel in Section \ref{sec5}.

%**********************************************************
%                                                         *
\subsection{Quantum Gaussian state}\label{toqgs section}% *
%                                                         *
%**********************************************************
%To consider a quantum Gaussian channel, 
We introduce 
a Gaussian density operator, which is defined to be 
a density operator with a characteristic function of the form, 
\begin{equation}\label{chara fuct}
{\Tr}\rho V(z)=\exp\left[im^tz-\frac{1}{2}z^t\alpha z \right],
\end{equation}
where $z$ is a column $2s$-dimensional vector  
$[x_1,...,x_s;y_1,..,y_s]^t$ and 
\begin{equation}
V(z)=\exp i\sum_{j=1}^s[x_jq_j+y_jp_j].
\end{equation}
In the characteristic function, 
$m$ is a column $2s$-vector and $\alpha$ is a real symmetric $2s\times2s$ 
matrix, satisfying 
\begin{equation}
m=\Tr \rho R,\quad \alpha-\frac{i}{2}\Delta=\Tr R \rho R^t,
\end{equation}
where $R=[q_1,..,q_s;p_1,...,p_s]$ and 
\begin{equation}
\Delta=
\left[
\begin{array}{cc}
O &\hbar I\\
-\hbar I& O
\end{array}
\right],
\end{equation}
with identity matrix $I$ and zero matrix $O$.
%********************************************************************
%                                                                   *
\subsection{Calculation of capacity}%*
\label{sec4.2}
%                                                                   *
%********************************************************************
\label{section qgc}
The mapping from classical parameter to quantum Gaussian state is,
in mathematical paper, called quantum Gaussian channel, 
which forms an important class of the continuous classical-quantum
channel described in Section \ref{ias section}.
Holevo and coworkers provided the general formula of capacity of 
this channel.
Take as input alphabet $\A$ the complex plane $\C$, 
and as the density operator $\rho (0)$ a Gaussian one with mean $0$.
Then the quantum Gaussian channel is described by the mapping
$\mu \to \rho(\mu)$,
where $\rho(\mu)=\D(\mu)\rho(0)$ with the displacement operator $\D(\mu)$.
%We can consider constraint on an input or an output of the mapping 
%$\mu \to \rho(\mu)$. 
Here we restrict ourselves to 
the case where we impose the input constraint.
The input constraint is given by 
putting $f(\mu)=\hbar\omega|\mu|^2$ in (\ref{ic0}) and (\ref{ic1}).

To carry out calculation of capacity for such a channel,
the following two properties of quantum Gaussian channel 
are essentially used \cite{Holevo:99}:\newline
{\bf (i)} the optimum {\it a priori} distribution $\pi$ in (\ref{cont capacity sub})
is  Gaussian, and the mixture 
$\rho_\pi =\int \rho(\mu) \pi (d\mu)$ is Gaussian again.
Let correlation matrices of {\it a priori} distribution $\pi$
and density operators $\rho$ and $\rho_\pi$ be 
$\beta$, $\alpha$ and $\gamma$ respectively.
Then the relation $\gamma=\alpha+\beta$ holds between them.
\newline
{\bf (ii) } von Neumann entropy of Gaussian density operator $\rho$
with a correlation matrix $\alpha$ is calculated as
\begin{equation}
H(\rho)=\frac{1}{2}{\rm Sp}G(-(\Delta^{-1}\alpha)^2),
\end{equation}
where $G(x^2)=(x+1/2)\log (x+1/2)-(x-1/2)\log (x-1/2)$.
In particular, for one mode Gaussian state the von Neumann entropy
is given by
\begin{equation}
H(\rho)=G\left( \frac{\alpha^{qq}\alpha^{pp}-(\alpha^{qp})^2}
{\hbar^2} \right).
\end{equation}
Thus the capacity
of the quantum Gaussian channel can be written as 
\begin{equation}\label{form1}
C=\max_{\beta\in B_1}\frac{1}{2}{\rm Sp} 
G(-[\Delta^{-1}(\alpha+\beta)]^2)
-\frac{1}{2}{\rm Sp}G(-(\Delta^{-1}\alpha)^2),
\end{equation}
where $B_1$ is the convex set of real positive matrices $\beta$,
satisfying
\begin{equation}
{\rm Sp}\varepsilon \beta\leq E,
\end{equation}
where 
\begin{equation}
\varepsilon
=
\left[
\begin{array}{cc}
\varepsilon_{1,1}&O\\
O&1/2I
\end{array}
\right],
\end{equation}
with 
$\varepsilon_{1,1}={\rm diag}[\omega_1^2/2,...,\omega_s^2/2]$, 
and $s\times s$ zero and identity matrices $O$, $I$.

In the formula (\ref{form1}) the maximization with respect to 
{\it a priori} distribution $\beta$ is left unsolved.
Holevo showed  \cite{Holevo:99} the explicit calculation of 
 (\ref{form1}) for one mode quantum Gaussian channel with 
{\it input} constraint.
The results are 
\newline
{\bf [A]} If the inequality 
\begin{equation}\label{ineq cond}
\left[ \frac{1}{2}\left( \omega \alpha^{qq}-\frac{\alpha^{pp}}{\omega}\right) \right]^2+(\alpha^{qp})^2\leq (\frac{E}{\omega})^2
\end{equation}
holds,
then we have
\begin{equation}
\begin{split}\label{cap1}
C=&G\left(\frac{1}{\hbar^2\omega^2}\left[ E+
\frac{1}{2}(\omega^2\alpha^{qq}
+\alpha^{pp})
\right]^2\right )\\
&-G\left(\frac{\alpha^{qq}\alpha^{pp}-(\alpha^{qp})^2}{\hbar^2} \right),
\end{split}
\end{equation}
where $G(d^2)=(d+1/2)\log (d+1/2)-(d-1/2)\log (d-1/2)$.
By putting $N_s=E/\hbar \omega$ and $N=\Tr \rho (0) \Qa^{\dagger}\Qa=(\omega^2\alpha^{qq}
+\alpha^{pp})/2\hbar\omega-1/2$, the first term in (\ref{cap1}) can be
written as
\begin{equation}
(N+N_s+1)\log(N+N_s+1)-(N+N_s)\log(N+N_s).
\end{equation}
When the inequality (\ref{ineq cond}) holds, 
the mixture $\rho_{\pi}=\int\rho(\mu)\pi(d\mu)$ with the optimum 
{\it a priori} distribution $\pi$ is always equal to the 
Bose-Einstein distribution with mean number of quanta $N+N_s$.
Moreover, assuming that $\rho(0)$ is a coherent state, we can 
also simplify the second term in Eq. (\ref{cap1})
and obtain the capacity as
\begin{equation}\label{cap_coh}
\begin{split}
C=&(N+N_s+1)\log(N+N_s+1)-(N+N_s)\log(N+N_s)\\
&-(N+1) \log (N+1) + N \log N\\
=&\log\left( 1+\frac{N_s}{N+1} \right)+
(N+N_s)\log \left( 1+\frac{1}{N+N_s}\right)\\
&-N\log \left(1+\frac{1}{N}\right).
\end{split}
\end{equation}
\newline
{\bf [B]} If the inequality (\ref{ineq cond}) does not hold,
we have 
\begin{equation}
\begin{split}\label{cap2}
C=&G\left(\frac{1}{\hbar^2\omega^2}\left\{\left[ E+\frac{1}{2}(\omega^2\alpha^{qq}
+\alpha^{pp})\right]^2\right . \right . \\
&\left . \left . -\left[ \sqrt{(\omega^2\alpha^{qq}-\alpha^{pp})^2/4+\omega^2(\alpha^{qp})^2}-E
 \right]^2\right \} \right )\\
&-G\left(\frac{\alpha^{qq}\alpha^{pp}-(\alpha^{qp})^2}{\hbar^2} \right).
\end{split}
\end{equation}

Note that the capacity under {\it output} constraint can be obtained 
by some modifications of the above discussion  (see \cite{Holevo:99}).

%******************************************************
%                                                     *
\section{Application to attenuated noisy channel}    %*
\label{sec5}
%                                                     *
%******************************************************
 The attenuated noisy channel corresponds to optical fiber channel or
space channel in optical communication system.  In this section, we show the
capacity formulae when  the communication process has such an attenuated
noisy channel. 
%In this case, it is shown that the capacity
%is not given by only classical-quantum channel  model.
\subsection{Gaussian state}
In this subsection we calculate the capacity 
%(\ref{form1})
for the attenuated noisy channel formulated in Sec \ref{qmooc section}. 
%which has already been  formulated in Section \ref{qmooc section}.
As a first step, we make some preparations for applying the results 
in Section  \ref{sec4.2} to this calculation.
\newline
{\bf (i) construction of the mapping $\mu\to\rho(\mu)$}
\newline
Since  squeezed state is the most general Gaussian state, we treat squeezed
state as an example of Gaussian state. 
The squeezed state $\tilde{\rho}(0)=\S(\gamma)|0\rangle$ is described as 
a pure Gaussian state with correlation matrix
\begin{equation}
\tilde{\alpha}=
\left[
\begin{array}{cc}
\tilde{\alpha}^{qq}&\tilde{\alpha}^{qp}\\
\tilde{\alpha}^{qp}&\tilde{\alpha}^{pp}
\end{array}
\right],
\end{equation}
with elements
\begin{eqnarray}
\tilde{\alpha}^{qq}&=&
\frac{\hbar}{2\omega}[\cosh 2\gamma-\sinh 2\gamma \cos \theta]\\
\tilde{\alpha}^{pp}&=&
\frac{\hbar\omega}{2}[\cosh 2\gamma +\sinh 2\gamma \cos \theta]\\
\tilde{\alpha}^{qp}&=&\frac{\hbar}{2}\sinh 2\gamma \sin \theta.
\end{eqnarray}
Then the output state $\rho(\mu)=T[\tilde{\rho}(\mu)]$ has the characteristic function \cite{Holevo:99b}
\begin{equation}
\Tr {\rho}(\mu) V(z)=\Tr \tilde{\rho}(\mu) V(kz)\cdot\exp \left[ -\frac{\hbar}{2}
\lambda(k,N_c) z^tz\right],
\end{equation}
where 
\begin{equation}
\lambda(k,N_c)=\left( \frac{1-k^2}{2}+N_c \right).
\end{equation}
This indicates that $\rho(\mu)=\D(k \mu)\rho'_0 \D(k \mu )^{\dagger}$, 
where $\rho'_0$ is the Gaussian state with the correlation matrix 
$\alpha=k^2\tilde{\alpha}+\hbar\lambda(k,N_c)\Qi$ and the mean 0.
\newline
{\bf (ii) transmitter constraint}
\newline
As shown in Section \ref{a2},
instead of input or output constraint given in Section \ref{section qgc},
we need to introduce 
a somewhat different constraint for reasonable evaluation
of the effect of squeezing.
%We should consider constraint on the average energy for 
%generating signal states at the transmitter.
That is, 
we impose a {\it transmitter} constraint
\begin{equation}
\Sp \varepsilon (\tilde{\alpha}+\beta)\leq \hbar\omega
\left(N_{tr}+\frac{1}{2}\right),
\end{equation}
where $\beta$ is a correlation matrix of
{\it a priori} probability distribution.
%Then, as shown in Figure \ref{ch model}, we need the model of attenuated noisy %channel
%rather than the general formulation of classical-quantum channel.

Thus the capacity with the attenuation channel can be written as
\begin{equation}
C=\max_{\beta\in B'_1}\frac{1}{2}\Sp 
G(-[\Delta^{-1}(\alpha+\beta)]^2)
-\frac{1}{2}\Sp G(-(\Delta^{-1}\alpha)^2),
\end{equation}
where $B'_1$ is the convex set of real positive matrices $\beta$,
satisfying
\begin{equation}
{\rm Sp}\varepsilon \beta\leq k^2 \left[\hbar\omega\left(N_{tr}+\frac{1}{2}\right)-
\Sp \varepsilon \tilde{\alpha}\right ].
\end{equation} 
We can obtain the explicit formula of the capacity
by replacing the energy bound $E$ in (\ref{ineq cond}), (\ref{cap1}) 
and (\ref{cap2})
with $k^2[\hbar\omega\left(N_{tr}+1/2\right)-
\Sp \varepsilon \tilde{\alpha}]$.
In the following we calculate the value of the capacity 
when $\omega=1$, $\tilde{\alpha}^{qp}=0$,
$\tilde{\alpha}^{qq}\geq \tilde{\alpha}^{pp}$  and $\tilde{\alpha}^{qq}\tilde{\alpha}^{pp}=\hbar^2/4$ hold.
By these substitutions, the inequality (\ref{ineq cond}) becomes
\begin{equation}\label{ineq cond2}% see 00-16 p.19 
\max\{ \tilde{\alpha}^{qq},\tilde{\alpha}^{pp} \}\leq 
\hbar (N_{tr}+1/2),
\end{equation}
and 
we have
\begin{equation}
\begin{split}
E+\Sp \varepsilon \alpha =&\hbar k^2 (N_{tr}+1/2)+\hbar \lambda(k,N_c)\\
=&\hbar\left(k^2N_{tr}+N_c+\frac{1}{2}\right).
\end{split}
\end{equation}
and
\begin{equation}
\begin{split}
\alpha^{qq}\alpha^{pp}=&(k^2\tilde{\alpha}^{qq}+\hbar\lambda(k,N_c))
                       (k^2\tilde{\alpha}^{pp}+\hbar\lambda(k,N_c))\\
                      =&\hbar^2[\frac{ k^4}{4}+k^2\lambda(k,N_c)(2N_{sq}+1)
                      +\lambda(k,N_c)^2]\\
                      =&\hbar^2\left[\left( N_c+\frac{1}{2} \right)^2 +
                      k^2 N_{sq}[(1-k^2)+2N_c]\right],
\end{split}
\end{equation}
where $N_{sq}$ is mean number of quanta for a transmitted 
squeezed state $\tilde{\rho}(0)$,
$N_{sq}=\Tr\tilde{\rho}(0)\Qa^{\dagger}\Qa$.
From this  the capacity is calculated as follows.
\newline
{\bf [A]} If the inequality (\ref{ineq cond2}) holds,
we obtain
\begin{equation}\label{cap1form}
\begin{split}
C=&(k^2N_{tr}+N_c+1)\log(k^2N_{tr}+N_c+1)\\
&-(k^2N_{tr}+N_c)\log(k^2N_{tr}+N_c)\\
%&G([k^2 (N_{tr}+1/2) +\lambda(k,N_c)]^2)\\
%&-G( k^4/{4}+k^2\lambda(k,N_c)(2N_{sq}+1)  +\lambda(k,N_c)^2)
&-G\left(\left[ N_c+\frac{1}{2} \right]^2 +
                      k^2 N_{sq}[(1-k^2)+2N_c]   \right).
\end{split}
\end{equation}
When we transmit coherent states, that is $N_{sq}=0$, 
the second term in (\ref{cap1form}) is simplified 
and the capacity is given by
\begin{equation}\label{cap2form}
\begin{split}
C=&(k^2N_{tr}+N_c+1)\log(k^2N_{tr}+N_c+1)\\
&-(k^2N_{tr}+N_c)\log(k^2N_{tr}+N_c)\\
&-(N_c+1)\log (N_c+1) +N_c \log N_c.
\end{split}
\end{equation}
As $G$ in (\ref{cap1form}) is a monotonously increasing function of $d^2$,
we can find that the capacity given by (\ref{cap1form}) is maximized when 
there is no squeezing $N_{sq}=0$ or 
when there is no noise and attenuation ($k=1$ and $N_c=0$).
That is, squeezing necessarily decreases the capacity 
of attenuated noisy channel.
\newline
{\bf[B]}
When the inequality (\ref{ineq cond2}) does not hold,
using
\begin{equation}
\begin{split}
|\alpha^{qq}-\alpha^{pp}|&/2-E \\
=&k^2|\tilde{\alpha}^{qq}-\tilde{\alpha}^{pp}|/2 
-\hbar k^2 (N_{tr}+1/2)+k^2 \Sp\tilde{\alpha}/2\\
=&k^2\max\{\tilde{\alpha}^{qq},\tilde{\alpha}^{pp} \}
-\hbar k^2(N_{tr}+1/2),
\end{split}
\end{equation}
we obtain
\begin{equation}
\label{cap3form}
\begin{split}
C=&G\left (k^2[2N_{tr}+1]\left[\lambda(k,N_c)+k^2 
\frac{ \max\{\tilde{\alpha}^{qq},\tilde{\alpha}^{pp}\}}{\hbar}\right]\right . \\
&\quad \quad\quad \left . +\lambda(k,N_c)^2
-k^4\frac{\max\{\tilde{\alpha}^{qq},\tilde{\alpha}^{pp}\}^2}{\hbar^2} \right )\\
%&-G\left( \frac{k^4}{4}+k^2\lambda(k,N_c)(2N_{sq}+1)
%                      +\lambda(k,N_c)^2\right).
&-G\left(\left[ N_c+\frac{1}{2} \right]^2 +
                      k^2 N_{sq}[(1-k^2)+2N_c]   \right).
\end{split}
\end{equation}
We would like to call these equations (\ref{cap1form}) and (\ref{cap3form}) {\it "quantum Shannon formula"} based on Holevo Theorem,
because these correspond to classical Shannon formula $C=\log(1+S/N)$.
\begin{figure}[t]　
\center
\includegraphics[scale=0.85,angle=0]{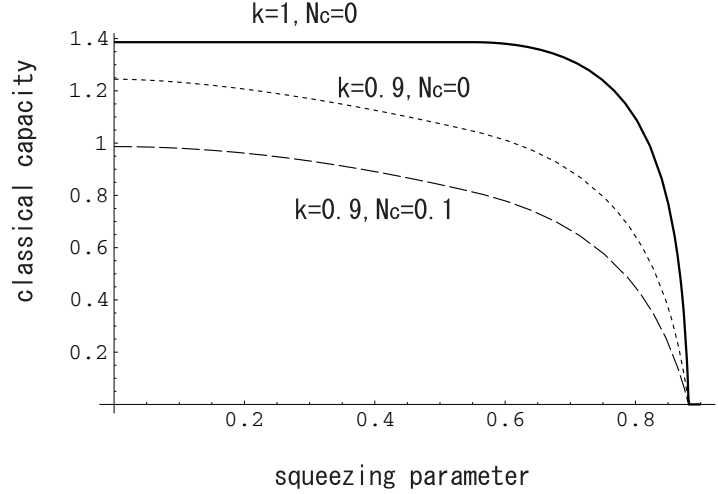}
\caption{capacity for the attenuated noisy channel}\label{gr_cap1}
\end{figure}
In Figure \ref{gr_cap1} we present graphs of the capacity with respect to 
squeezing parameter $\gamma$, when $(k,N_c)=(1,0)$, $(k,N_c)=(0.9,0)$ and 
$(k,N_c)=(0.9,0.1)$.
These graphs show that 
the capacity for the non-attenuated noiseless channel 
does not change if squeezing is not too large, 
while that for attenuated noisy channel is necessarily decreased 
by any squeezing.

%Furthermore we shall consider the case where the coherent amplitude of the squeezing 
%state is restricted to real number i.e. 
%\begin{equation}
%\beta^{pp}=0, \quad \beta^{qp}=0
%\end{equation}
%Let us suppose that the equality of (\ref{ineq cond2}) holds
%without loss of generality.
%Then  we have
%\begin{equation}
%\beta^{qq}=k^2[\hbar(2N_{tr}+1)-(\tilde{\alpha}^{qq}+\tilde{\alpha}^{pp})],
%\end{equation}
%and hence the capacity is calculated as 
%\begin{equation}
%\begin{split}
%C=&G\left( [-k^2 \frac{\tilde{\alpha}^{pp}}{\hbar}+\lambda(k,N_c)+k^2(2N_{tr}+1%)][k^2\fra%c{\tilde{\alpha}^{pp}}{\hbar}+\lambda(k,N_c)] \right)\\
%&-G\left( \frac{k^4}{4}+k^2\lambda(k,N_c)(2N_{sq}+1)
%                      +\lambda(k,N_c)^2\right).
%C=&G\left( \left[ -k^2 \frac{\tilde{\alpha}^{pp}}{\hbar} +
%\frac{1+k^2}{2}+N_c+2k^2N_{tr}\right]
%\left[ k^2\frac{\tilde{\alpha}^{pp}}{\hbar}+\frac{1-k^2}{2} \right] \right)\\%01-1,48
%&-G\left(\left[ N_c+\frac{1}{2} \right]^2 +
%                      k^2 N_{sq}[(1-k^2)+2N_c]   \right).
%\end{split}
%\end{equation}
%figure\begin{figure}[t]　
%\center
%\includegraphics[scale=0.6,angle=0]{capFig3.eps}
%\caption{capacity for the attenuated noisy channel 
%in the case that the coherent amplitude is restricted to real number}\label{gr_cap2}
%\end{figure}
%In Figure \ref{gr_cap2} we present graphs of the capacity with respect to 
%squeezing parameter $\gamma$, when $(k,N_c)=(1,0)$, $(k,N_c)=(0.9,0)$ and 
%$(k,N_c)=(0.9,0.1)$.
\subsection{Number state}
If we employ photon number state as the transmitter state, then alphabets
are
discrete number : $n$. The density operator is described by
\begin{equation}
\rho = \sum_n P(n) | n \ra\la n |
\end{equation}
In the case of no attenuation process, the maximum entropy is given when
\begin{equation}
P(n) = \frac{1}{1+<n>}\left(\frac{<n>}{1+ <n>}\right )^n
\end{equation}
which is called Bose-Einstein distribution. $<n>$ is average photon number.
So the capacity becomes
\begin{equation}
C = \log(1+<n>) + <n>\log(1 + \frac{1}{<n>})
\end{equation}
For such a photon signal, the channel models of attenuation and
amplification processes were discussed by Shimoda, Takahashi, and Townes\cite{Shimo:57}.
The channel model of the noiseless attenuation process is described by
binomial distribution as follows:
\begin{equation}
P(n_y | n_x ) = \frac{{n_x}!}{{n_y}!(n_x -n_y )!} k^{n_y} (1 - k)^{n_x -
n_y}
\end{equation}
where $n_x$ and $n_y$ are input and output photon numbers, respectively.
This means that if the input state is certain number state $ | n_x \ra\la n_x
|$,
then the output is described by
\begin{equation}
\rho_{n_x} = \sum_{n_y}\frac{{n_x}!}{{n_y}!(n_x -n_y )!} k^{n_y} (1 -
k)^{n_x - n_y}
| n_y \ra\la n_y |
\end{equation}
If the probability distribution in the input is $P(n_x)$,
then the output density operator becomes
\begin{equation}
\rho_{out} = \sum_{n_y} \sum_{n_x} P(n_x) P(n_y | n_x ) | n_y \ra\la n_y|
\end{equation}
So the channel capacity is
\begin{equation}
C = \max_{P(n_x)} \{H(\rho_{out}) - \sum P(n_x) H(\rho_{n_x})\}
\end{equation}
We have no solution in this case. However, if we assume that
the input distribution is Bose-Einstein distribution,
the output distribution is also Bose-Einstein.
Then, the Holevo-Yuen-Ozawa bound\cite{Yuen-Ozawa:93} 
bound becomes for $<n_y> \gg 1$
\begin{equation}
\Delta H = \frac{1}{2} (\log <n_y> - \log(2\pi) + 1 + \gamma)
\end{equation}
for $<n_y> \ll 1$
\begin{equation}
\Delta H = (1-\gamma)<n_y>
\end{equation}
where $\gamma$ is Euler constant(0.5772...).

\section{Information theoretical meaning of ultimate channel capacity--
Binary discretization}
In this section we turn our attention to the idea of {\it discretization}
\cite{Sohma:00}, which
is introduced by the equation (\ref{discrete support})
to treat the continuous channel analytically.
The discretization means deriving discrete channels from the original 
continuous channel by restricting the number of letters used in 
the information transmission to a finite one.
The properties of the original continuous channel can be determined 
by the behavior of all such derived discrete channels.
The main purpose of this section is to show that
 the  discretization for the quantum continuous channel
has properties  inherent in the quantum system.
To this purpose, we recall the Gordon's suggestion\cite{Gordon:62} that the 
binary quantum counter can extract essentially all the information 
incorporated in a weak light wave. Basing on this suggestion, 
we infer that the {\it binary discretization},
restricting the number of letters 
to only two, realizes asymptotically the capacity 
in the quantum case, while
 the binary discretization necessarily causes some loss of information 
 in the classical case.
In the following we shall verify this inference by investigating 
the binary discretization of the noiseless coherent state channel.
We shall calculate the capacity given by the optimum binary discretization,
and compare it with the capacity of the original continuous channel.
Further we shall also consider the binary discretization for
the maximum mutual information; this binary discretization 
is closely related with 
the binary model presented in the Gordon's suggestion.
%This allows us to clarify the meaning of Gordon's suggestion 
%in the light of moder quantum information theory.

Now let us demonstrate the binary discretization of the coherent state channel,
input signals are 
coherent states $\{ |\alpha \rangle \}$ ($\alpha \in {\C}$) 
and the corresponding {\it a priori} distribution $\pi$ 
 is  constrained by
\begin{eqnarray}
\int_{\C} |\alpha |^2 \pi (d^2 \alpha) \le m.\label{ICC}
\end{eqnarray}
The coherent state channel is
a basic example of the quantum Gaussian channel,
and its capacity is given by putting $N_s=m$ and $N=0$ in (\ref{cap_coh})
as,
\begin{eqnarray}
C_{BE}=(m+1)\log (m+1) - m\log m. \label{Gordon entropy}
\end{eqnarray}
%This quantity was first anticipated by Gordon as an upper bound for 
%the amount of information transmitted by a light wave, 
%and later given rigorously according to the definition of the capacity
%of the quantum continuous channel \cite{Holevo:98}.

Let us find the optimum binary discretization
 for the capacity of the coherent state channel, 
 solving the optimization problem:
\begin{equation}
C^{(2)}=\sup_{\{\alpha,\beta\}} \sup_{Q} 
H(\{ \alpha , \beta \}, \{ Q, 1-Q \}).\label {binary capacity}
\end{equation}
Here the suprema are taken over all binary set of inputs, 
$\{\alpha ,\beta \}$,  
 and all probability assignments $Q, 1-Q$ 
 satisfying the constraint
\begin{eqnarray}
Q |\alpha|^2+(1-Q)|\beta|^2 \le m, \label{binary constraint}
\end{eqnarray} 
and
 $H(\{ \alpha , \beta \}, \{ Q, 1-Q \})$ is the capacity of 
binary channel with input signals $\{ | \alpha \rangle , | \beta \rangle \}$
and the corresponding {\it a priori} probabilities $\{ Q, 1-Q \}$.

We can calculate the quantity $C^{(2)}$ as
%\begin{multline}
\begin{equation}
C^{(2)}=-\left [ \frac{1-e^{-2m}}{2}\log( \frac{1-e^{-2m}}{2}) 
+ \frac{1+e^{-2m}}{2}\log( \frac{1+e^{-2m}}{2})\right ]. \label{C2}
%\end{multline}
\end{equation}
Here the optimum binary discretization is given by the symmetric binary signals 
 $\{ |\alpha \rangle , |-\alpha \rangle \}.$
\begin{figure}[t]　
\includegraphics[scale=0.75,angle=0]{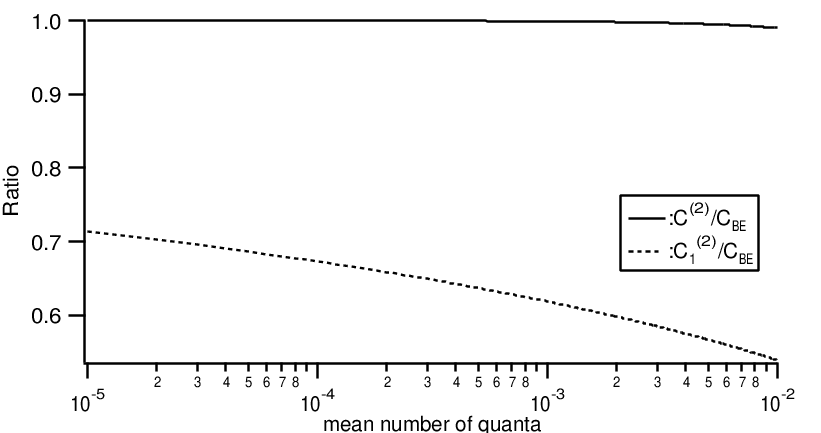}
\caption{$C^{(2)}/C_{BE}$ and $C^{(2)}_1/C_{BE}$ with respect to mean number of quanta $m$
}\label{cap fig}
\end{figure}
In Figure \ref{cap fig}
 the value of $C^{(2)}/C_{BE}$ is plotted 
with respect to 
the energy constraint $m$.
The figure shows that the binary discretization realizes 
approximately the capacity $C_{BE}$ in a weak photon case ($ m < 10^{-2}$).
Indeed, by applying $e^{-x}\approx 1-x$ and $\log(1-x)\approx -x$ 
to Eqs. (\ref{Gordon entropy}) and (\ref{C2}) and neglecting the term of $m^2$,
the following approximation holds,
\begin{eqnarray}
C_{BE}\approx C^{(2)} \approx -m \log m + m \quad \quad \mbox{for }m \ll 1.
\end{eqnarray}
These results show that
the coherent state channel can be simulated by the binary 
discrete channel in a weak photon case.

On the other hand, the classical continuous channel
can not be simulated by any discrete channel\cite{Cover:91}.
In particular the binary discretization does not realize the capacity.
To achieve the capacity, 
we must solve a more complicated optimization problem.
The problem, 
related with 
the {\it sphere packing}, 
is still one of main topics in the classical information theory.

We shall consider another optimization, 
which is concerned with
the maximum mutual information and formulated as follows,
%\begin{multline}
\begin{equation}
C_1^{(2)}=\sup_{\{\alpha, \beta \}}\sup_{Q} \sup_{\{\hat{\Pi}_{\alpha},\hat{\Pi}_{\beta}\}}%\\
I_d(\{\alpha ,\beta \},\{Q,1-Q\} , \{\hat{\Pi}_{\alpha},\hat{\Pi}_{\beta}\}).
\label{C1n}
\end{equation}
%\end{multline}
Here $I_d$ is the mutual information of the binary channel 
with input signals $\{|\alpha\rangle,|\beta\rangle \}$, the 
corresponding {\it a priori} probabilities $\{Q, 1-Q\}$ and 
the signal detection process 
given by the positive operator-valued measure (POVM)
$\{\hat{\Pi}_{\alpha},\hat{\Pi}_{\beta}\}$.
The optimum POVM for two input letters consists of two detection
operators \cite{Davies:78}, and hence we can restrict 
the number of the detection operators 
in Eq. (\ref{C1n}) to two without loss of generality.

Let us solve the optimization problem.
The right-hand side of Eq. (\ref{C1n}) can be divided into two parts
as follows,
\begin{gather}
C_1^{(2)}= \sup_{\{ \alpha  ,\beta \}} \sup_{Q} 
I_a(\{ \alpha ,\beta \},Q)\label{C1n bin}\\
I_a(\{ \alpha  ,\beta \},Q)=
\sup_{\{\hat{\Pi}_1,\hat{\Pi}_2\}}I_d(\{ \alpha  ,\beta  \},\{Q,1-Q\},\{\hat{\Pi}_1,\hat{\Pi}_2\}).\label{I_a}
\end{gather}
The  accessible information is 
 known \cite{Ban:99} to be calculated as
\begin{eqnarray}
I_a=I_a(\{\alpha ,\beta  \},Q)=H(Q)-H(f(\{ \alpha ,\beta \},Q)).\label{def Ia}
\end{eqnarray}
In this equation 
\begin{gather}
H(Q)=-Q\log Q -(1-Q)\log (1-Q),\\
f=f(\{ \alpha ,\beta \},Q)=\frac{1}{2}\left ( 1-\sqrt{1-4\kappa ^2Q(1-Q)} \right ),
\end{gather}
where
\begin{equation}
\kappa^2 = |\langle \alpha | \beta \rangle |^2 = \exp[-|\alpha -\beta|^2].
\end{equation}

On the contrary to $C^{(2)}$, the quantity $C^{(2)}_1$ can be only 
obtained approximately:
the optimum {\it a priori} distribution is given by 
\begin{eqnarray}
Q_{opt}(m)\approx \frac{-m\log m -m}{3},\label{approx Q}
\end{eqnarray}
and the optimum binary signals are 
$\{|\alpha\ra,|\beta\ra\}$ satisfying $\beta=k \alpha$ $(-1\le k \le 0)$ and 
\begin{subequations}
\begin{eqnarray}
|\alpha |^2&=&\frac{1-Q_{opt}(m)}{Q_{opt}(m)}m,\\
|\beta |^2&=&\frac{Q_{opt}(m)}{1-Q_{opt}(m)}m.
\end{eqnarray}
\end{subequations}
Then The value of $C_1^{(2)}$ is given by  $I_a(Q_{opt}(m))$.

In Fig \ref{cap fig}
  the value of $C_1^{(2)}/C_{BE}$ is plotted 
with respect to the energy constraint $m$.
The ratio $C_1^{(2)}/C_{BE}$ converges to 1 when $m \to 0$.
This substantiates the validity of the Gordon's suggestion in theory.
However the convergence speed is very slow.
For example $C_1^{(2)}/C_{BE}$ is 0.82 for $m=10^{-10}$, 
and then the capacity takes a very small value: $C_{BE} \approx 2.4\times 10^{-9}$.
This shows the Gordon's suggestion is not correct in practice.
We can further obtain the following approximation of $C_1^{(2)}$, 
the value of which is larger than the capacity $\log(m+1)$
 but is less than the capacities $C^{(2)}$, $C_{BE}$:
 \begin{equation}
 \begin{split}
 \log(m+1)&\le C_1^{(2)}\approx -m\log m - m \\
 &\le C^{(2)}\approx C_{BE}\approx -m \log m + m \\
 &\quad \quad \quad \mbox{ for } m\ll 1.
 \end{split}
 \end{equation}
Strictly speaking, $C_1^{(2)}$ should be compared to the maximum 
mutual information $C_1$ of the original quantum 
continuous channel. It is conjectured $C_1^{(2)}$ gives
 a good approximation of $C_1$, but we have no way of 
 calculating $C_1$ for the quantum continuous channel.

\section{Conclusion}
In this review paper, applications of the general formulae for  capacity  in
the quantum information theory to optical communication processes have been
introduced. As a special result, we would like to emphasize that capacity
formula was calculated for attenuated noisy channel by using Holevo's
general theory of capacity for continuous alphabet. This formula may provide
the capacity formula in optical field which corresponds to Shannon capacity
formula for Gaussian noise model in long distance  microwave communication
system.

%\bibliography{mybibs}

\begin{thebibliography}{10}

\bibitem{Ban:99}
M.~Ban, K.~Kurokawa, and O.~Hirota.
\newblock Cut-off rate performance of quantum communication channels with
  symmetric signal states.
\newblock {\em J. of Opt. B, Quantum Semiclass. Opt.}, 1:206--218, 1999.

\bibitem{Ban:97}
M.~Ban, K.~Kurokawa, R.~Momose, and O.~Hirota.
\newblock Optimum measurements for discrimination among symmetric quantum
  states and parameter estimation.
\newblock {\em Int. J. Theor. Phys}, 36:1269--1288, 1997.

\bibitem{Bennett:98}
C.~H. Bennett and P.~W. Shor.
\newblock Quantum information theory.
\newblock {\em IEEE Trans. on Information Theory}, 44(6):2724--2742, 1998.

\bibitem{Davies:78}
Davies.
\newblock Information and quantum measurement.
\newblock {\em IEEE. Trans. Inform. Theory}, 24:596--599, 1978.

\bibitem{Fuchs:94}
C.~A. Fuchs and C.~M. Caves.
\newblock Ensemble-dependent bounds for accessible information in quantum
  mechanics.
\newblock {\em Phys. Rev. Lett.}, 73:3047--3050, 1994.

\bibitem{Fuchs:96}
C.~A. Fuchs and A.~Peres.
\newblock Quantum state disturbance versus information gain: Uncertainty
  relations for quantum information.
\newblock {\em Phys. Rev. A}, 53:2038--2045, 1996.

\bibitem{Gordon:62}
J.~P. Gordon.
\newblock Quantum effect in communications systems.
\newblock {\em IRE Proc.}, 50:1898--1908, 1962.

\bibitem{Haus:96}
P.~Hausladen, R.~Joza, B.~Schumacher, M.~Westmoreland, and W.~Wootters.
\newblock {\em Phys. Rev. A}, 54:1869, 1996.

\bibitem{Hirota:00}
O.~Hirota.
\newblock A foundation of quantum channels with super additiveness for shannon
  information.
\newblock {\em Applicable Algebra in Eng. Communication and Computing},
  10(4/5):401--427, 2000.

\bibitem{Holevo:73}
A.~S. Holevo.
\newblock Statistical detection theory for quantum systems.
\newblock {\em J. of Multivariable analysis}, 3:337--394, 1973.

\bibitem{Holevo:98}
A.~S. Holevo.
\newblock The capacity of quantum communication channel with general signal
  states.
\newblock {\em IEEE Trans. Inform. Theory}, 44:269--273, 1998.

\bibitem{Holevo:98b}
A.~S. Holevo.
\newblock Coding theorems for quantum channels.
\newblock {\em Tamagawa University Research Review}, 4(1), 1998.

\bibitem{Holevo:99}
A.~S. Holevo, M.~Sohma, and O.~Hirota.
\newblock Capacity of quantum gaussian channels.
\newblock {\em Phys. Rev. A}, 59:1820--1828, 1999.

\bibitem{Holevo:99b}
A.~S. Holevo and R.~F. Werner.
\newblock Evaluating capacities of bosonic gaussian channels.
\newblock {\em Phys. Rev. A}, 63:032312, 2001.

\bibitem{Kato:99b}
K.~Kato, M.~Osaki, and O.Hirota.
\newblock Derivation of classical capacity of quantum channel for discrete
  information source.
\newblock {\em Physics Letters A}, 251:157--163, 1999.

\bibitem{Levitin:95}
L.~B. Levitin.
\newblock Optimal quantum measurements for two pure and mixed states.
\newblock {\em Plenum Press, New York, ed by Belavkin, Hirota, and Hudson},
  1995.

\bibitem{Osaki:98}
M.~Osaki, M.~Ban, and O.~Hirota.
\newblock {\em J. of Modern Optics}, 45:269--282, 1998.

\bibitem{Sasaki:98}
M.~Sasaki, K.~Kato, M.~Izutu, and O.~Hirota.
\newblock Quantum channels showing superadditivity in classical capacity.
\newblock {\em Phys. Rev. A}, 58:146--158, 1998.

\bibitem{Schu:97}
B.~Schumacher and M.~D. Westmoreland.
\newblock Sending classical information via noisy quantum channel.
\newblock {\em Phys. Rev. A}, 56(1):131--138, 1997.

\bibitem{Shimo:57}
K.~Shimoda, H.~Takahashi, and C.H.Townes.
\newblock Fluctuations in amplification of quanta with application to maser
  amplifiers.
\newblock {\em J. Phys. Soc. Japan}, 12(6):686, 1957.

\bibitem{Sohma:00}
M.~Sohma and O.~Hirota.
\newblock Binary discretization for quantum continuous channels.
\newblock {\em Physical Review A}, 62(5):052312--1--4, 2000.

\bibitem{Yuen-Ozawa:93}
H.~P. Yuen and M.~Ozawa.
\newblock Ultimate information carrying limit of quantum systems.
\newblock {\em Phys. Rev. Lett.}, 70:363, 1993.

\end{thebibliography}

\end{document}